\documentclass[useAMS,usenatbib]{mn2e}
\usepackage[dvips]{graphicx}
\usepackage{times}

\newcommand{\aap}{A\&A}
\newcommand{\mnras}{MNRAS}
\newcommand{\pasp}{PASP}
\newcommand{\apj}{ApJ}
\newcommand{\apjl}{ApJ}
\newcommand{\pasj}{PASJ}
\newcommand{\aj}{AJ}

\title[Comparative study of $N$-body and FP simulations - II]
      {Comparative study between $N$-body and Fokker-Planck simulations for rotating star clusters
       - II. 2-component models}
\author[J. Hong, E. Kim, H. M. Lee, \& R. Spurzem]
{Jongsuk Hong$^1$\thanks{E-mail: jshong@astro.snu.ac.kr (JH); eunhyeuk@kari.re.kr (EK); hmlee@snu.ac.kr (HML); spurzem@bao.ac.cn (RS)}
 , Eunhyeuk Kim$^2$\footnotemark[1], Hyung Mok Lee$^{1,3}$\footnotemark[1], \& Rainer Spurzem$^{4,5}$\footnotemark[1] \\
$^1$Astronomy Program, Department of Physics and Astronomy, Seoul National University , 1 Gwanak-ro, Gwanak-gu, Seoul 151-742, Korea\\
$^2$Korea Aerospace Research Institute, 169-84 Gwahak-ro, Yuseong-gu, Daejeon 305-806, Korea\\
$^3$Center for Theoretical Physics, Seoul National University , 1 Gwanak-ro, Gwanak-gu, Seoul 151-742, Korea\\
$^4$National Astronomical Observatories, Chinese Academy of Sciences 20A Datun Road, Chaoyang District, Beijing, China\\
$^5$Astronomisches Rechen-Institut, Zentrum fur Astronomie Univ. Heidelberg (ZAH) Monchhofstrasse 12-14, 69120 Heidelberg, Germany}

\begin{document}

\date{Accepted 2013 January 14. Received 2012 November 28}

\pagerange{\pageref{firstpage}--\pageref{lastpage}} \pubyear{2013}

\maketitle

\label{firstpage}
\begin{abstract}
To understand the effects of the initial rotation on the evolution of the tidally limited clusters with mass spectrum,
we have performed $N$-body simulations of the clusters with different initial rotations 
and compared the results with those of the Fokker-Planck (FP) simulations. 
We confirmed that the cluster evolution is accelerated by not only the initial rotation but also the mass spectrum.
For the slowly rotating models, the time evolutions of mass, energy and angular momentum show good agreements between $N$-body and FP 
simulations. 
On the other hand, for the rapidly rotating models, there are significant differences between these two approaches at the early stage of the evolutions because of the development of bar instability in $N$-body simulations. 
The shape of the cluster for $N$-body simulations becomes tri-axial or even prolate, 
which cannot be produced by the 2-dimensional FP simulations. 
The total angular momentum and the total mass of the cluster decrease rapidly while bar-like structure persists. 
After the rotational energy becomes smaller than the critical value for the bar instability, 
the shape of the cluster becomes nearly axisymmetric again, and follows the evolutionary track predicted by the FP equation.
We have confirmed again that the energy equipartiton is not completely achieved when \(M_{2}/M_{1}(m_{2}/m_{1})^{3/2}>\) 0.16.
By examining the angular momentum at each mass component, we found that the exchange of angular momentum between different mass components occurs, 
similar to the energy exchange leading to the equipartition.  
\end{abstract}


\section{Introduction}
The effects of initial rotation on the dynamical evolution of star clusters received substantial attention
because the rotation is a natural consequence during the formation process. 
The current population of star clusters may not show significant amount of rotation, 
but it does not mean that the initial conditions inhibit the presence of rotation. 
The direct measurement of the rotation is difficult since it requires long integration with large aperture telescope. 
Rather indirect evidence for the rotation comes from the shape of globular clusters. 
Although most globular clusters show high degree of circular symmetry, 
the ellipticity has been measured for large number of star clusters \citep[e.g.,][]{1987ApJ...317..246W,2010ApJ...721.1790C}. 
If the ellipticity is due to the rotational flattening, many clusters still have some degrees of rotation. 
Even though the amount of rotation in current population of globular clusters is rather small,
initial clusters could have been rotating much more rapidly 
since the rotation phases out as the clusters evolve dynamically.

The effects of rotation on the dynamical evolution have been studied by a number of authors. 
\citet{1983PhDT.........4G} has extended  the Fokker-Planck (FP) equation for rotating systems, 
but his study was limited to slowly rotating systems by imposing the spherical symmetry for the shape of the clusters. 
The thermodynamical analyses have been pioneered by \citet{1979PASJ...31..523H, 1982PASJ...34..313H} 
and found that there exists an instability similar to gravothermal catastrophe 
and they named this phenomenon as `gravo-gyro catastrophe'. 
Theses earlier studies provided the basis of the possible acceleration of dynamical evolution due to the initial rotation.

More careful studies for the rotating systems have been carried out by 
\citet{2002MNRAS.334..310K, 2004MNRAS.351..220K, 2008MNRAS.383....2K}
using 2-Dimensional FP code developed by \citet{1999MNRAS.302...81E}.
These papers investigated both isolated and tidally limited clusters, and single and multi-component clusters. 
The general result emerged from these studies is that the rotating clusters undergo faster evolution 
than non-rotating ones for single component models. 
The acceleration is also expected in multi-mass models as well, but the degree of acceleration could be significantly reduced 
since the energy exchange between different mass components is another accelerating process and these two processes compete each other. 

The suitability of the FP approach to the study of dynamical evolution of star clusters 
has been a matter of debate because the absence of the accurate knowledge on the third integral 
for rotating systems prohibits us to include all the possible integrals in constructing the FP equation. 
Comparison with $N$-body calculation should provide a clue to the validity of the current version of the FP approach 
\citep[e.g.,][]{1994MNRAS.270..298G,1994MNRAS.268..257G,1994MNRAS.269..241G}. 
Such a comparison for rotating systems was done by \citet{2007MNRAS.377..465E} and \citet{2008MNRAS.383....2K} for single component models 
and showed that the FP results are generally consistent with the $N$-body calculations.

We extend the comparison to the two-component models as an interim step to the full multi-mass models. 
Two-component models have the ingredients for the multi-mass models, but have smaller model parameters. 
The important difference between single and multi-mass models is the existence of energy exchange among 
different mass components. In rotating models, there is also a possibility of exchange of the specific 
angular momentum through the dynamical process. It is much easier to investigate such processes in $N$-body.
Another motivation for carrying out $N$-body simulations and comparing with FP results is 
to address the validity of the axisymmetric assumption which is inevitable for the FP approaches. 
If the cluster rotates rather rapidly, the bar-like structure can form even with the initial assumption of the axisymmetric shape. 
The evolution of the elongated cluster could be different from the perfectly axisymmetric one. 

This paper is organized as follows: In Section 2, we describe the models and their parameters in detail. 
The effect of the initial rotation on the cluster shape will be presented in Section 3. 
In Section 4, we will compare $N$-body and FP results in various angles.
We will discuss the effects of mass spectra on the dynamical evolution of star clusters in Section 5.

\section{The Models}
Most of FP results for rotating stellar system with initial mass spectrum
are based on the 2D FP solver, mFOPAX \citep{2004MNRAS.351..220K} which is the revised version of
FOPAX \citep{1999MNRAS.302...81E, 2002MNRAS.334..310K} suitable for the rotating stellar
system with initial mass spectrum. For the complete description of mFOPAX
readers are referred sections 2 and 3 of \citet{2004MNRAS.351..220K}.

The \textsc{nbody} code which we used for this study is one of series of direct $N$-body programs developed by S. Aarseth since 1960s.
Each version of the \textsc{nbody} codes has been added some epochal schemes such as the Ahmad-Cohen neighbor scheme,
the Kustaaheimo-Stiefel or chain regularization scheme \citep{1999PASP..111.1333A}.
More recently, the \textsc{nbody4} and the \textsc{nbody6} codes can perform more precise calculations thanks to the 4th-order Hermite integrator \citep{1999PASP..111.1333A}. 
Specially, the \textsc{nbody4} code is designed for running on the GRAPE, 
which is a special-purpose machine for only direct $N$-body simulations by calculating gravity or coulomb interaction with high parallelization.
We used the GRAPE6-BLX64 boards for the $N$-body calculations reported in this paper. 

To prepare the initial models for the present $N$-body runs we have assigned the positions and the velocities of the stars
from the predefined density, potential and distribution function of initial models. These initial models also
used in FP runs are obtained following \citet{1987AJ.....93.1106L} and also applied
in previous studies \citep{1999MNRAS.302...81E, 2002MNRAS.334..310K, 2004MNRAS.351..220K, 2008MNRAS.383....2K}.
 Globular clusters or open clusters which are typical stellar systems considered
in the present study are tidally limited by their host galaxy.
To investigate the time evolution of rotating stellar system under
the influence of the tidal effect of the host galaxy and to compare the
dynamical evolution of rotational stellar system using two different
numerical methods (FP and $N$-body), we assume that the stellar systems are
orbiting around the center of the Galaxy with circular orbit, which is already
applied in previous FP runs \citep{2002MNRAS.334..310K, 2004MNRAS.351..220K}. 
In order to directly compare $N$-body results with FP methods, we remove the stars whose total energy exceeding
the tidal energy due to the host galaxy instantaneously \citep[i.e., energy cut-off,]
[]{1997MNRAS.292..331T,2001MNRAS.325.1323B,kim_phd,2007MNRAS.377..465E,2008MNRAS.383....2K}. 
According to the $N$-body computation, it takes at least a crossing time to escape from the clusters 
for stars with total energy larger than the tidal energy.
Therefore, instantaneous removal of stars with energy greater than tidal energy is somewhat unrealistic. 
The problem becomes more serious for small-$N$ systems since the fraction of stars to be unbounded 
at a given time is higher than large-$N$ systems \citep{1998ApJ...503L..49T, 2000MNRAS.318..753F}.
However, since the main goal of the present study is to investigate the difference between $N$-body and FP methods for rotating
stellar systems,
we need to apply the same criteria with the FP approach. 
The tidal boundary (or tidal energy) is adjusted during evolution as described 
in the previous studies \citep{2002MNRAS.334..310K, 2004MNRAS.351..220K, 2008MNRAS.383....2K}.

\begin{table*}
  \begin{center}
  \caption{Initial parameters for all models.}
  \begin{tabular}{c c c c c c c c c c c}
    \hline
    \hline
    Model & \(W_{0}\) &  \(\omega_{0}\) &  \(r_{tid}/r_{c}\) & \(T_{rot}/|W|\) & \(N\)*runs &  \(m_{2}/m_{1}\) & \(M_{1}/M_{2}\) & \(N_{2}\) & \(S\) & \(\Lambda\)\\    \hline
        &   & 0.0 &           18.0 & 0.000 & 20,000*3 & & & & &\\
        &   & 0.3 &           14.5 & 0.035 & 20,000*3 & & & & &\\
M2A$^\dagger$ & 6 & 0.6 & \phantom{0}9.9 & 0.101 & 20,000*3 &2&5&1818&0.566&1.056\\
        &   & 0.9 & \phantom{0}7.1 & 0.156 & 20,000*3 & & & & &\\
        &   & 1.2 & \phantom{0}5.4 & 0.196 & 20,000*3 & & & & &\\
        &   & 1.5 & \phantom{0}4.4 & 0.222 & 20,000*3 & & & & &\\
\\
        &   & 0.0 &           18.0 & 0.000 & 20,000*1 & & & & &\\
    M2B & 6 & 0.6 & \phantom{0}9.9 & 0.101 & 20,000*1 &5&5&\phantom{0}769&2.236&9.518\\
        &   & 1.2 & \phantom{0}5.4 & 0.196 & 20,000*1 & & & & &\\
\\
        &   & 0.0 &           18.0 & 0.000 & 20,000*1 & & & & &\\
    M2C & 6 & 0.6 & \phantom{0}9.9 & 0.101 & 20,000*1 &10&5&\phantom{0}392&6.325&50.24\\
        &   & 1.2 & \phantom{0}5.4 & 0.196 & 20,000*1 & & & & &\\
\\
        &   & 0.0 &           18.0 & 0.000 & 20,000*1 & & & & &\\
    M2D & 6 & 0.6 & \phantom{0}9.9 & 0.101 & 20,000*1 &20&5&\phantom{0}198&17.89&256.2\\
        &   & 1.2 & \phantom{0}5.4 & 0.196 & 20,000*1 & & & & &\\
\\
        &   & 0.0 &           18.0 & 0.000 & 20,000*1 & & & & &\\
    M2Ae& 6 & 0.6 & \phantom{0}9.9 & 0.101 & 20,000*1 &2&20&\phantom{0}488&0.141&0.264\\
        &   & 1.2 & \phantom{0}5.4 & 0.196 & 20,000*1 & & & & &\\
    \hline
    \multicolumn{11}{l}{$^\dagger$Fokker-Planck simulations are performed for comparison.}
  \end{tabular}
  \end{center}
  \label{tbl1}
\end{table*}
Table \ref{tbl1} shows the initial parameters used for the $N$-body simulations. 
For comparison, we also performed FP simulations of M2A models.
There are several parameters that determine the cluster evolutions such as the concentration parameter \(W_{0}\),
the initial rotation \(\omega_{0}\) \citep{1999MNRAS.302...81E, 2002MNRAS.334..310K,2007MNRAS.377..465E}, the mass spectrum \citep{2004MNRAS.351..220K,2007MNRAS.374..703K}
and tidal boundary \citep{2001MNRAS.325.1323B,kim_phd,2007MNRAS.377..465E}.
We fixed \(W_{0}=6\), but varied \(\omega_{0}\) and mass spectrum to investigate the effect of 
the initial rotation and the mass spectrum on the cluster evolution.
The initial rotation \(\omega_{0}\) are varied from 0.0 to 1.5.
For the mass spectrum, we varied the individual mass ratio \(m_{2}/m_{1}\) from 2 to 20 
while the total mass ratio \(M_{1}/M_{2}\) is fixed to 5 or 20. 
One of the most important processes of the stellar system with multiple mass components is the energy equipartition.
The energy equipartition is a tendency for different mass components to have similar kinetic energies.
However, in some cases, the equipartition is not completely achieved, which is called as the equipartition instability \citep{1969ApJ...158L.139S}.
To determine whether the equipartition happens to be achieved or not in two-component mass systems, \citet{1969ApJ...158L.139S} derived an analytic equipartition stability parameter,
\begin{equation}
S=\frac{M_{2}}{M_{1}} \bigg( \frac{m_{2}}{m_{1}} \bigg)^{3/2}.
\end{equation}
He suggested that the energy equipartition between low and high mass stars takes place when \(S < S_{crit} = 0.16\).
After Spitzer's study, many authors have studied the energy equipartition process of two-component systems 
by theoretical approaches \citep[e.g.,][]{1978ApJ...221..567L} and 
by several numerical methods such as
Monte-Carlo approaches to solve the FP equation \citep{1971ApJ...164..399S}, the direct integration of 
FP equation \citep{1998ApJ...495..786K} and $N$-body simulations \citep{2000ApJ...528L..17P}.
\citet{2000ApJ...539..331W} performed Monte-Carlo simulations with various two-component mass spectra and introduced an empirical equipartition stability parameter
\begin{equation}
\Lambda=\frac{M_{2}}{M_{1}}\bigg( \frac{m_{2}}{m_{1}} \bigg)^{2.4}
\end{equation}
and found that the critical value for energy equipartition is \(\Lambda_{crit}=\) 0.32.
Most of our models have \(S > S_{crit}\) and \(\Lambda > \Lambda_{crit}\).
The ratio of the rotational kinetic energy to the potential energy is known to be a measure of the `temperature' 
of the rotating system with higher value being called cold system. 
If this parameter is greater than 0.14, the system is known to become dynamically unstable against the formation of the bar-like structure  
\citep{1973ApJ...186..467O}.
Thus, models with \(\omega_{0}\) greater than 0.9 are expected to evolve to elongated shape.
We designate these models as rapidly rotating models and 
the other models with \(\omega_{0}\le0.6\) as slowly rotating models in this study.

\section{Slowly and Rapidly Rotating Clusters}
To investigate the effect of the initial rotation on the evolution of clusters, we focus on M2A models which have various amounts of the initial rotation.
\begin{figure}
  \centering
  \includegraphics[width=84mm]{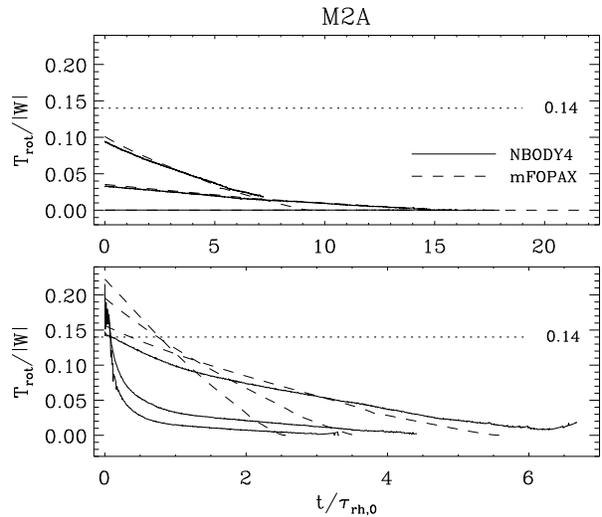}
  \caption{The ratio of rotational kinetic energy to potential energy. Solid and dashed lines represent $N$-body and FP results, respectively.
  Upper panel shows the ratios for slowly rotating models and lower for those of the rapidly rotating models.
  The dotted lines denote a criterion of bar instability \citep{1973ApJ...186..467O}. \
  For rapidly rotating models the $N$-body and FP results show the significant difference due to the bar instability. }
  \label{fig_ins}
\end{figure}
In Fig. \ref{fig_ins}, we show the time evolution of the ratio of the rotational kinetic energy to the potential energy, \(T_{rot}/|W|\).
The dashed and solid lines represent FP and $N$-body results, respectively.
The initial half-mass relaxation time is measured as follows suggested by \citet{1971ApJ...164..399S},
\begin{equation}
\tau_{rh,0}=0.138\frac{N^{1/2}r_{h,0}^{3/2}}{G^{1/2}\overline{m}^{1/2}\ln\Lambda},
\end{equation}
where \(N,r_{h,0},G,\overline{m}\) and \(ln\Lambda\) are total number of stars, initial half-mass radius,
gravitational constant, mean mass of stars and Coulomb logarithm, respectively.
It is well known that a system with rigid-body rotation suffers a secular instability 
when \(T_{rot}/|W|\) is larger than 0.14 \citep{1973ApJ...186..467O}.
Later, from numerical simulations, \citet{1981A&A....99..362S} confirmed that the criterion is valid for more realistic rotation curves.
Our models with initial value of \(T_{rot}/|W| < 0.14\) are shown in the upper panel and 
those with \(T_{rot}/|W| > 0.14\) are shown in the lower panel. 
The results of $N$-body and FP show similar behaviors for the models with initial \(T_{rot}/|W| < 0.14\).
However, for the models with initial \(T_{rot}/|W| > 0.14\), the initial evolution depends on different numerical approaches.
The $N$-body simulation represents more rapid decrease of \(T_{rot}/|W|\) with time than the FP simulation in the early phase. 
For the rapidly rotating models one can observe the construction of bar-like structure and the total rotational energy decreases very quickly.
Therefore, the FP approach seems to be not appropriate in describing the evolution of rotating models with initial value of \(T_{rot}/|W|>0.14\). 

\begin{figure}
  \centering
  \includegraphics[width=84mm]{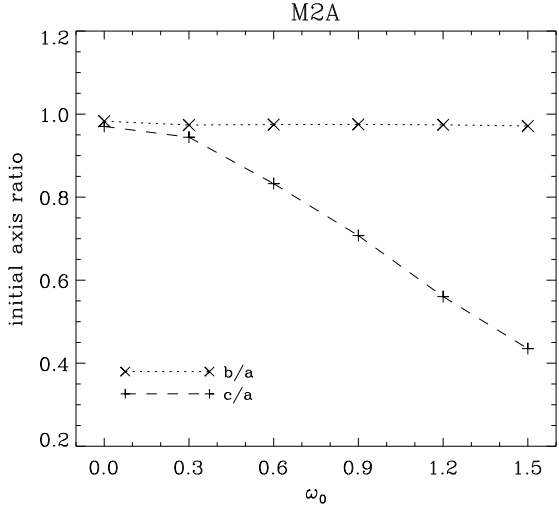}
  \caption{ Initial axis ratio of clusters with different initial rotations. Cross and plus symbols show intermediate (\(b/a\)) and minor (\(c/a\))
  axis ratios, respectively.
  We estimate axis ratios with stars in half mass ellipsoidal radius.
  The clusters show more oblate shapes as the initial rotation increases. }
  \label{fig_ax0}
\end{figure}
In order to investigate the evolution of the cluster shape,
we calculate axis ratios of clusters by using the method suggested by \citet{1991ApJ...378..496D}.
They defined a tensor,
\begin{equation}
M_{ij}=\Sigma\frac{x_{j}x_{j}}{q^{2}}
\end{equation}
with an ellipsoidal radius, 
\begin{equation}
q=\biggl( x^{2}+\frac{y^{2}}{(b/a)^{2}}+\frac{z^{2}}{(c/a)^{2}} \biggr)^{1/2},
\end{equation}
where \(a, b\) and \(c\) are axis lengths with \(a \ge b \ge c\).
The axis ratios are derived from the tensor through
\begin{equation}
\frac{b}{a}=\biggl(\frac{M_{yy}}{M_{xx}}\biggr)^{1/2}\quad\textrm{and}\quad
\frac{c}{a}=\biggl(\frac{M_{zz}}{M_{xx}}\biggr)^{1/2},
\end{equation}
where \(M_{xx}, M_{yy}\) and \(M_{zz}\) are the principal components of the tensor.
In order to compute the tensor \(M_{ij}\), we need to know the axis ratios. 
Therefore, for the simultaneous determination of the tensor and axis ratios, we need to perform an iterative calculation. 
We, first, assume certain set of axis ratios (e.g., \(b/a = 1\) and \(c/a = 1\)) and compute \(M_{ij}\), which gives another set of axis ratios. 
Obviously, the resulting axis ratios will be different from the assumed values, and therefore, can be used as an input for improved estimation of \(M_{ij}\). 
We carry out the iteration until the relative difference of axis ratios becomes less than certain criterion (we include a value of \(10^{-4}\) for this study).
Fig. \ref{fig_ax0}  shows the axis ratios as a function of initial rotation for M2A models.
We calculate axis ratios with stars in the ellipsoidal radius including half-mass of the cluster.
The shape of rotating cluster is oblate initially due to the initial rotation (i.e., \(b/a = 1\) and \(c/a < 1\)).
The minor axis ratio \(c/a\) decreases when the initial rotation increases from 0.3 to 1.5. 

\begin{figure}
  \centering
  \includegraphics[width=84mm]{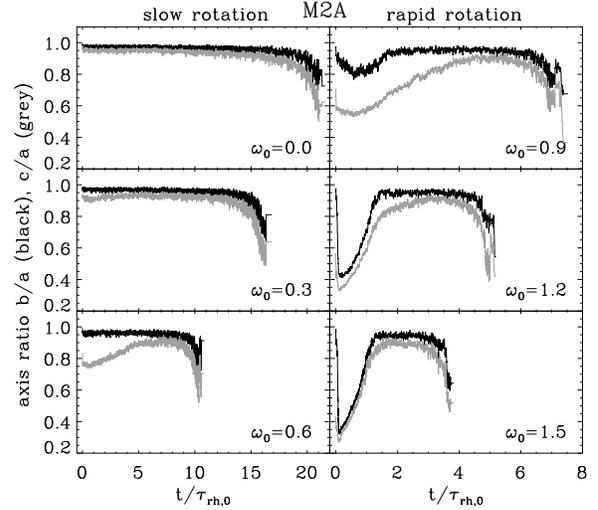}
  \caption{ Time evolution of intermediate \(b/a\) (black) and minor \(c/a\) (grey) axis ratios for M2A models.
  In the early stage for rapidly rotating models, the intermediate axis ratios decrease rapidly due to the bar instability.
  The shapes of clusters tend to become spherical after a few  \(\tau_{rh,0}\) due to the significant loss of angular momentum.}
  \label{fig_axr}
\end{figure}
In Fig. \ref{fig_axr}, we show the evolution of the axis ratios, \(b/a\) and  \(c/a\) for M2A models with different initial rotations.
The left panels are the result of slowly rotating models and right panels are for the rapidly rotating models.
The minor axis ratios \(c/a\) increases with time for slowly rotating models because the cluster loses angular momentum.
For the rapidly rotating models, the intermediate axis ratio decreases at the beginning,
because of the development of the bar instability.
Due to this instability, cluster shapes become tri-axial or even prolate in a dynamical time scale which is much shorter than the relaxation time.
The intermediate axis ratio also decreases during this phase.
The decrease of both axis ratios is more rapid when the initial rotation becomes larger.
When the ratio \(T_{rot}/|W|\) becomes smaller than 0.14 by the loss of the rotational energy as shown in Fig. \ref{fig_ins},
the bar instability disappears and the cluster becomes axisymmetric.

\begin{figure*}
  \includegraphics[width=150mm]{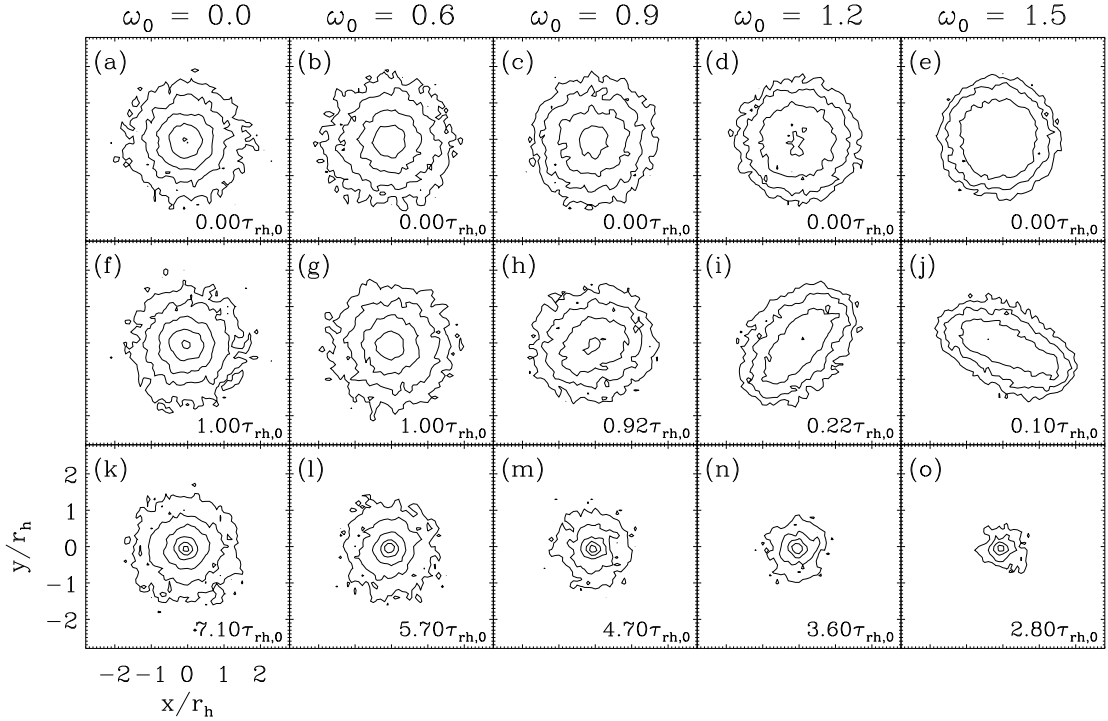}
  \caption{ Projected density contour maps of M2A models with initial rotation \(\omega_{0} = 0.0, 0.6, 0.9, 1.2\) and 1.5 on XY plane.
  The top panels ($a$-$e$) show initial models and bottom panels ($k$-$o$) show density structures at the time of the core collapse.
  The middle panels of \(\omega_{0} = 0.9, 1.2, 1.5\) ($h$-$j$) are density structures at the time of maximum elongation. }
  \label{fig_cnt}
\end{figure*}
Density contour maps on XY plane of clusters with different initial rotation parameters are shown in Fig. \ref{fig_cnt}.
The top and bottom panels show the initial shapes and the shapes at the core collapse, respectively. 
In the bottom panels, the size of contours show that the remaining mass of cluster at the core collapse becomes smaller
as the initial rotation increases.
Density structures for models with \(\omega_{0} = 0.9, 1.2\) and 1.5 in middle panels represent shapes
at the time of maximum elongation.
One can clearly observe bar-like structures of rapidly rotating models.
Shapes become to be those of the axisymmertic systems at the time of core collapse as shown in Fig \ref{fig_axr} 
(i.e., \(b/a=1\) and \(c/a<1\)).

\section{Comparison between $N$-body and FP Results}

\subsection{Mass and Energy}
\begin{figure}
  \centering
  \includegraphics[width=84mm]{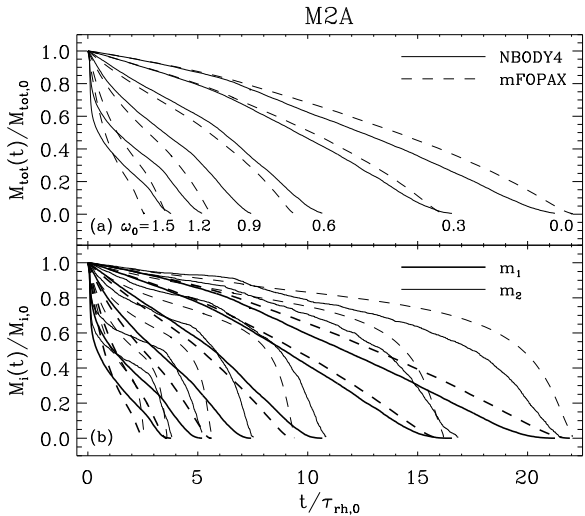}
  \caption{ Evolution of mass for M2A models: (a) Evolution of total mass. Dashed lines represent FP results, and solid lines show $N$-body results.
   The total mass decreases due to escaping stars through the tidal energy threshold.
   For rapidly rotating models, there are steep mass losses at the early stages because of the bar instability.
   (b) Time evolution of the mass components \(m_{1}\) (red) and \(m_{2}\) (blue).
   Total mass of \(m_{1}\) decreases faster than that of \(m_{2}\).}
  \label{fig_mss}
\end{figure}
In this section, we compare the evolution of $N$-body and FP simulations for M2A models with three different point of views: 
overall evolution (mass and energy), central evolution (central density and velocity dispersion) and rotational evolution.
Fig. \ref{fig_mss}a shows the evolutions of total mass.
For the slowly rotating models, the time evolutions of total cluster mass for $N$-body simulations agree well
with those of FP results.
However, there exist significant differences between FP and $N$-body results
for the rapidly rotating models.
For an instance, $N$-body calculations show significantly higher mass loss rate than FP results for rapidly rotating models 
with \(\omega_{0} =\) 1.2 and 1.5, especially in the very early times. 
The significant amount of mass loss at the very early stage also induces the large amount of angular momentum loss.
After this stage, clusters have smaller number of stars with slower rotation and
therefore evolve slowly compared to FP results.
The time evolutions of total mass of each component are shown in Fig. \ref{fig_mss}b.
Because the number of low mass stars is much larger than that of high mass stars,
the evolution of total mass is similar to that of low mass stars.

\begin{figure}
  \centering
  \includegraphics[width=84mm]{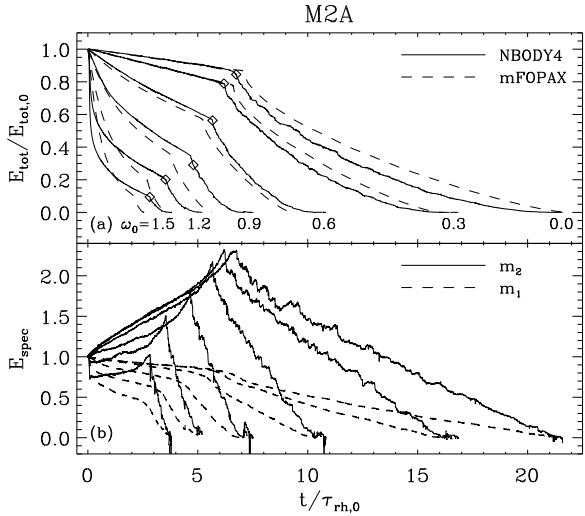}
  \caption{Upper panel shows the evolution of the total energy as a function of time.
  The diamond symbols represent the moment of the core collapse for $N$-body results.
  Total energies decrease with time due to escaping stars.
  After the core collapse, energies decrease more rapidly because of core bounce due to the binary heating.
  In lower panel, there are the evolutions of the specific energies for \(m_{1}\) and \(m_{2}\).
  The specific energy of \(m_{2}\) increases quickly during the just prior of the core collapse
  because high mass stars move to the central region having deep potential due to the mass segregation.}
  \label{fig_enr}
\end{figure}
The total energy of the tidally limited stellar system decreases monotonically with time
due to the escaping stars carrying energies.
Fig. \ref{fig_enr}a shows the time evolution of the normalized total energies.
The results of $N$-body and FP simulations agree well for slowly rotating models.
While the total energy decreases slowly in pre-collapse stage,
after core collapse the total energy decreases more rapidly due to the core expansion.
For the rapidly rotating models, however FP and $N$-body results show significant difference 
because of the large amount of energy loss during the early stages for $N$-body models.
Fig. \ref{fig_enr}b represents the time evolution of normalized specific energy (i.e., energy per unit mass) for each mass component.
In overall, the specific energies of \(m_{1}\) continue to decrease.
On the other hand, the specific energy of \(m_{2}\) increases slowly till the core collapse as a result of equipartition and mass segregation. 
However, the specific energies of \(m_{2}\) decrease after the core collapse due to the mass loss through the tidal boundary
 because the process of mass segregation stops at the core collapse 
(i.e. the mean mass of a star in the central region increases until the core collapse and remains as a constant value after the core collapse;
see \S 5.3).
Note that the mean energies for both low mass and high mass stars immediately decrease for the model of \(\omega_{0}=1.2\) and 1.5.
Again, this is caused by the bar instability in the early phase of the evolution.

\subsection{Central Density, Velocity Dispersion and Core Collapse}
\begin{figure}
  \centering
  \includegraphics[width=84mm]{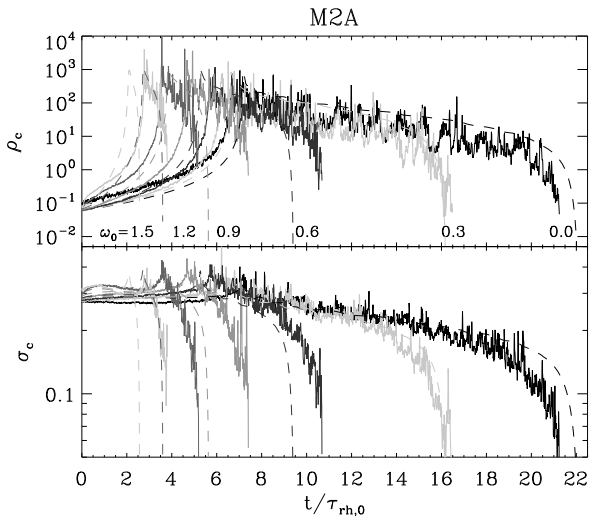}
  \caption{ Evolution of central density \(\rho_{c}\) and velocity dispersion \(\sigma_{c}\) for M2A models.
  Dashed lines represent the FP results and the solid lines for the $N$-body results.
  Different contrasts mean models with different initial rotations.
  \(\rho_{c}\) and \(\sigma_{c}\) of rapidly rotating models show significant differences between two approaches.}
  \label{fig_rv}
\end{figure}
Fig. \ref{fig_rv} shows time evolution of the central density, \(\rho_{c}\) and the central velocity dispersion, \(\sigma_{c}\)
obtained by using stars inside the core radius.
The dashed and solid lines represent FP and $N$-body results, respectively, and different contrasts represent 
models with different initial rotation.
We confirm that the rotation accelerates both the core collapse and cluster disruption \citep[e.g.,][]{2008MNRAS.383....2K}.
In addition, FP and $N$-body results agree well for slowly rotating models. 
However, as the initial rotation increases the difference between FP and $N$-body results becomes large due to the bar instability.
Table \ref{tbl2} lists the core collapse times and the disruption times of M2A models.
For example, the model with \(\omega_{0}=1.5\) reaches the core collapse at 2.1\(\tau_{rh,0}\) 
and 3.7\(\tau_{rh,0}\) for FP and $N$-body calculations, respectively.
As the initial rotation \(\omega_{0}\) increases, the cluster spends more time of its whole life in the pre-core collapse phase (i.e., \(t_{cc}/t_{dis}\) increases).
For $N$-body simulations, the ratio of the core collapse time to the disruption time \(t_{cc}/t_{dis}\) 
of the model with \(\omega_{0}=1.5\) is \(\sim\)3/4 while that of the model with \(\omega_{0}=0\) is \(\sim\)1/3.
We observe that the core collapse and disruption times of $N$-body results are significantly longer than 
those of FP results for rapidly rotating models because these models are redefined as small and slowly rotating systems after the bar instability.
Although there are significant differences between $N$-body and FP time scales,
the total mass at the time of core collapse (designated as \(M_{cc}\)) of $N$-body and FP results show a good agreement.
The presence of the mass spectrum also accelerates the evolution of cluster \citep{2004MNRAS.351..220K}. 
We found that the core collapse times of M2A models are 20\(-\)40\% smaller than those of single mass systems considered by \citet{2008MNRAS.383....2K}.
On the other hand, the disruption times are only 10\% smaller.
We also look into the evolution of clusters with various mass spectra.
Table \ref{tbl3} shows the collapse and disruption time scales for other $N$-body models.
As \(m_{2}/m_{1}\) increases, the evolution of clusters is accelerated.
From M2A to M2D models, the core collapse time decreases when \(m_{2}/m_{1}\) increases, but is less affected by the initial rotation.
The disruption time, however, depends on both the mass spectrum and the initial rotation.
\begin{table}
  \begin{center}
  \caption{Global evolution of M2A models for \textsc{nbody4} and mFOPAX results}
  \begin{tabular} {c c c c c c}
    \hline
    \hline
    Simul.&  \(\omega_{0}\) &  \(t_{cc}/\tau_{rh,0}\) & \(t_{dis}/\tau_{rh,0}\) & \(t_{cc}/t_{dis}\) &\(M_{cc}\) \\
    \hline
           & 0.0 & 7.1 & 22.0 & 0.32 & 0.80 \\
           & 0.3 & 6.7 & 15.9 & 0.42 &0.70 \\
    mFOPAX & 0.6 & 5.3 & \phantom{1}9.2 & 0.58 & 0.52 \\
           & 0.9 & 3.8 & \phantom{1}5.7 & 0.67 & 0.36 \\
           & 1.2 & 2.7 & \phantom{1}3.7 & 0.73 & 0.23 \\
           & 1.5 & 2.1 & \phantom{1}2.5 & 0.84 & 0.14 \\
    \hline    
           & 0.0 & 6.8 & 21.2 & 0.32 & 0.80 \\
           & 0.3 & 6.2 & 16.3 & 0.38 & 0.72 \\
    \textsc{nbody4} & 0.6 & 5.7 & 10.4 & 0.55 &0.54 \\
           & 0.9 & 4.8 & \phantom{1}7.1 & 0.68 & 0.35 \\
           & 1.2 & 3.6 & \phantom{1}5.2 & 0.69 & 0.23 \\
           & 1.5 & 2.8 & \phantom{1}3.7 & 0.76 & 0.15 \\
    \hline
  \end{tabular} 
  \label{tbl2}
  \end{center}
\end{table}
\begin{table}
  \begin{center}
  \caption{Global evolution of other \textsc{nbody4} models}
  \begin{tabular} {c c c c c c c}
    \hline
    \hline
    Model&  \(\omega_{0}\) &  \(t_{cc}/\tau_{rh,0}\) & \(t_{dis}/\tau_{rh,0}\) & \(t_{cc}/t_{dis}\) & \(M_{cc}\) \\
    \hline
              & 0.0 & 1.61 & 10.18 & 0.16 & 0.94 \\ 
           M2B& 0.6 & 1.64 & 5.01 & 0.33 & 0.79 \\
              & 1.2 & 1.38 & 2.16 & 0.64 & 0.34 \\
    \hline  
              & 0.0 & 0.59 & 5.92 & 0.10 & 0.97 \\
           M2C& 0.6 & 0.62 & 2.79 & 0.22 & 0.89 \\
              & 1.2 & 0.70 & 1.36 & 0.51 & 0.44 \\
    \hline  
              & 0.0 & 0.42 & 3.59 & 0.12 & 0.96 \\
           M2D& 0.6 & 0.34 & 1.65 & 0.21 & 0.91 \\
              & 1.2 & 0.36 & 0.72 & 0.5 & 0.55 \\
    \hline
              & 0.0 & 9.37 & 27.56 & 0.34 & 0.73 \\
          M2Ae& 0.6 & 6.72 & 12.86 & 0.52 & 0.49 \\
              & 1.2 & 4.05 & 5.89 & 0.69 & 0.22 \\
    \hline
  \end{tabular}
  \label{tbl3}
  \end{center}
\end{table}

\begin{figure}
  \centering
  \includegraphics[width=84mm]{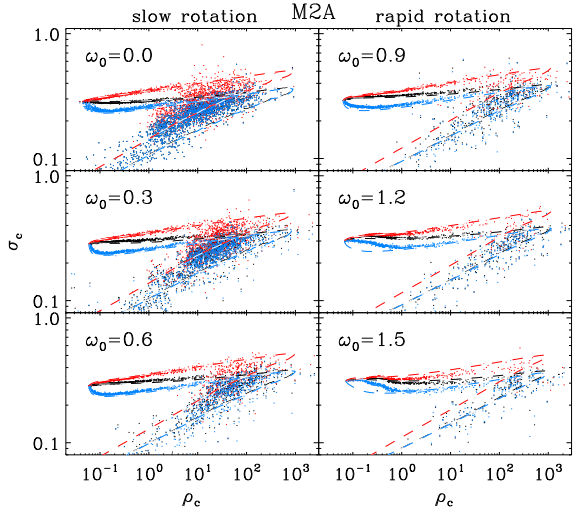}
  \caption{Central velocity dispersion \(\sigma_{c}\) for \(m_{1}\) (red), \(m_{2}\) (blue) and all stars (black) as a function of central density \(\rho_{c}\).
  Dots show $N$-body results, and dashed lines mean FP results. 
  Initially, \(\sigma_{c}\) for both mass components are the same. They evolve into the different ways due to the equipartition.
  \(\sigma_{c}\) for low mass becomes about \(\sqrt{2}\)-times larger than that for high mass because of equipartition.}
  \label{fig_sig}
\end{figure}

Fig. \ref{fig_sig} shows the evolution of the central velocity dispersion (\(\sigma_{c}\)) of each mass component as a function of \(\rho_{c}\).
\(\sigma_{c}\) of \(m_{1}\) and \(m_{2}\) are divided into two parts and increase gradually until the core collapse.
The total \(\sigma_{c}\) approaches that of \(m_{2}\) because the fraction of high mass stars in the core increases with time 
and the core is finally filled with high mass starts due to the mass segregation.
We can clearly see the equipartition as two distinct branches of \(\sigma_{c}\) versus \(\rho_{c}\).
As a result of equipartition, \(\sigma_{c}\) of \(m_{1}\) becomes about \(\sqrt{2}\)-times 
greater than that of \(m_{2}\) because we use individual mass ratio of \(m_{2}/m_{1} = 2\).
Once the establishment of energy equipartition, the evolutions of \(\sigma_{c}\) of \(m_{1}\) and \(m_{2}\) are represented by simple power-law.
During the post-core collapse stage, however, FP and $N$-body results show some differences.
$N$-body results are more dispersive than FP results, unlike the early evolution.
This is because there is the large fluctuation of \(\rho_{c}\) as shown in Fig. \ref{fig_rv}, which is from the lack of stars 
within the core radius after core collapse.
On the other hand, rapidly rotating models that lead to the bar instability have different evolutionary behaviors 
between FP and $N$-body for the very early phase. 
Because the bar instability delays the relaxation processes, the mass segregation in $N$-body takes place later compared to FP models. 
The gap between \(\sigma_{c}\) of \(m_{1}\) and \(m_{2}\) for $N$-body results is slightly smaller than that of FP results. 
This means that the bar instability also affects the equipartition process (see \S 5.2 and Table \ref{tbl4} for more details).
\begin{figure}
  \centering
  \includegraphics[width=84mm]{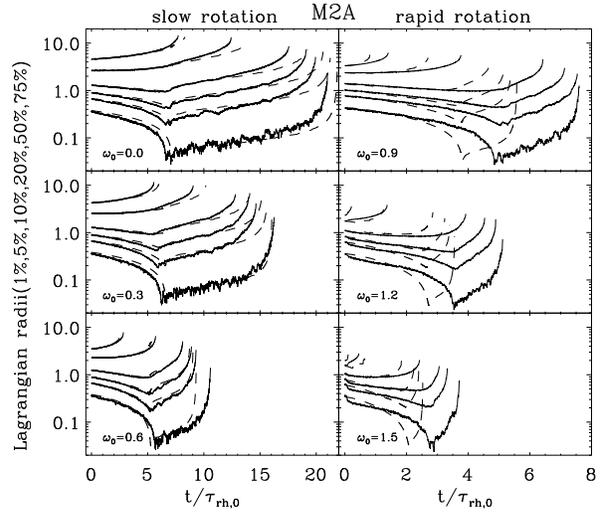}
  \caption{Lagrangian radii including 1, 5, 10, 20, 50, 75 per cent of initial cluster mass for M2A models.
  For rapidly rotating models, the results of $N$-body and FP shows significant differences
  while those for slowly rotating models are similar.}
  \label{fig_lgr}
\end{figure}

In Fig. \ref{fig_lgr}, we show the time evolution of the Lagrangian radii for M2A models.
We estimate Lagrangian radii containing 1, 5, 10, 20, 50 and 75 per cent of initial cluster mass.
The dashed lines and the solid lines represent the FP and $N$-body results, respectively.
Although our models are flattened due to the initial rotation and become triaxial by the bar instability, 
we use the spherical radius to estimate these Lagrangian radii for the simplicity of comparison between $N$-body and FP simulations.
We found that, for the slowly rotating models,
the results of FP and $N$-body show an excellent agreement except for the final stages when there are only small number of stars in the clusters.
However, for the rapidly rotating models, the results of FP and $N$-body are significantly different.
Again, the bar instability and induced mass loss are the main reason for these differences.

\subsection{Rotational properties}
\begin{figure}
  \centering
 \includegraphics[width=84mm]{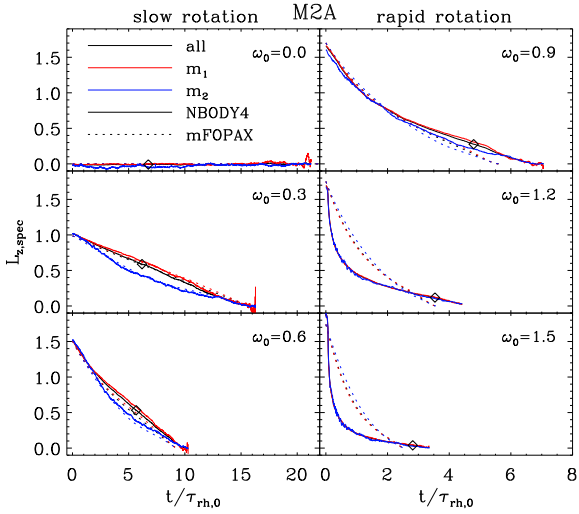}
  \caption{ Time evolution of the specific angular momentum \(L_{z,spec}\). Solid lines are the results of FP and
   dashed lines are the results of $N$-body. Red and blue colors mean low mass and high mass stars, respectively.
   Diamond symbols represent the moment of the core collapse. For slowly rotating models (left panels), \(L_{z,spec}\) of \(m_{1}\) and \(m_{2}\)
   are divided. However, \(L_{z,spec}\) of \(m_{1}\) and \(m_{2}\) decrease quickly compared to the results of FP
   for rapidly rotating models (right panels).}
  \label{fig_ang}
\end{figure}

To understand the effects of initial rotations on the cluster dynamics, 
we investigate the evolution of angular momentum.
Because we assume that stars escape through the tidal energy threshold, 
clusters lose their angular momentum continuously.
Also, we expect exchange of angular momentum between different mass components, similar to energy exchange.
After encounters, stars that lose their energy spiral into the central parts 
while stars that gain the energy move outward.
Also, loss of angular momentum makes the orbits of stars be eccentric
but gaining angular momentum does the orbits be less eccentric.
Fig. \ref{fig_ang} shows the time evolution of specific angular momentum which is defined as
\begin{equation}
\vec{L}_{spec}=\frac{\sum_{i}m_{i}\vec{r_{i}}\times\vec{v_{i}}}{\sum_{i}m_{i}}
\end{equation}
where \(\vec{r_{i}}\), \(\vec{v_{i}}\) are relative positions and velocities of stars to the center of mass, respectively,
and we integrate all stars in the cluster to compare $N$-body and FP results.
\(L_{z,spec}\) is z-direction component of \(\vec{L}_{spec}\).
Solid lines are the results of $N$-body and dashed lines represent the results of FP.
Red and blue colors show low and high mass stars, respectively.
For slowly rotating models, the results of $N$-body and FP are similar. 
The specific angular momenta decrease monotonically.
Due to the mass segregation, high mass stars that encounter with many low mass stars
migrate to the inner part where rotational velocity is smaller than that of the outer part of the cluster
(see rotation curves in Fig. \ref{fig_rot06}a and Fig. \ref{fig_rot15}a).
This is also related to the angular momentum exchange between different mass components (see \S 5.4 for more details). 
Thus, the specific angular momentum of \(m_{2}\) becomes smaller than that of \(m_{1}\) during the evolution.
The difference of angular momentum between two mass components is most prominent at the core collapse,
where the core collapse time is marked as diamond symbols in Fig. \ref{fig_ang}.
For the models with \(\omega_{0} = 1.2\) and \(1.5\), the angular momentum decreases rapidly during the early stage.
In addition, there is no clear split of the specific angular momentum of low and high mass stars 
due to the rapid evolution induced by the bar instability.
Interestingly, the model with \(\omega_{0} = 0.9\) shows the transitional evolution.
While the model still has a triaxial shape (\(T<1.5\tau_{rh,0}\), see Fig. \ref{fig_axr}), 
the evolution is similar to that of the models with \(\omega_{0} = 1.2\) and \(1.5\). 
However, after the shape becomes the axisymmetric again, 
the evolution is similar to that of the models with \(\omega_{0} = 0.3\) and \(0.6\). 

\subsubsection{Slow rotation (\(\omega_{0} = 0.6\))}
\begin{figure}
  \centering
  \includegraphics[width=84mm]{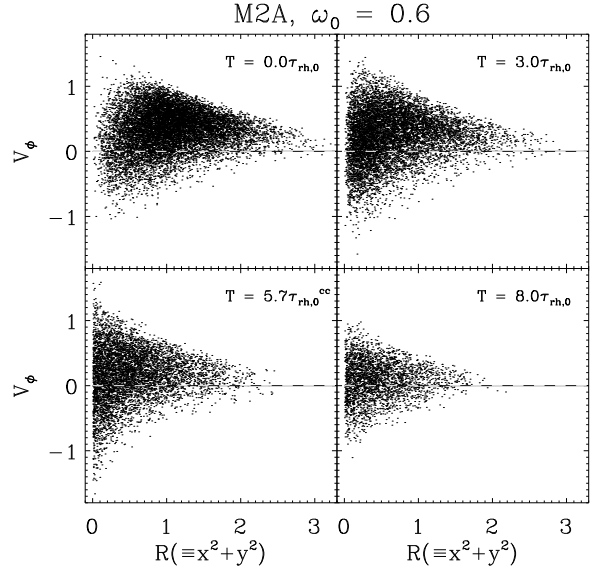}
  \caption{ Distribution of the rotational velocities of stars as a function of cylindrical radius.
  The cluster's size and degree of the anisotropy decrease with time.}
  \label{fig_tr06} 
\end{figure}
\begin{figure}
  \centering
  \includegraphics[width=84mm]{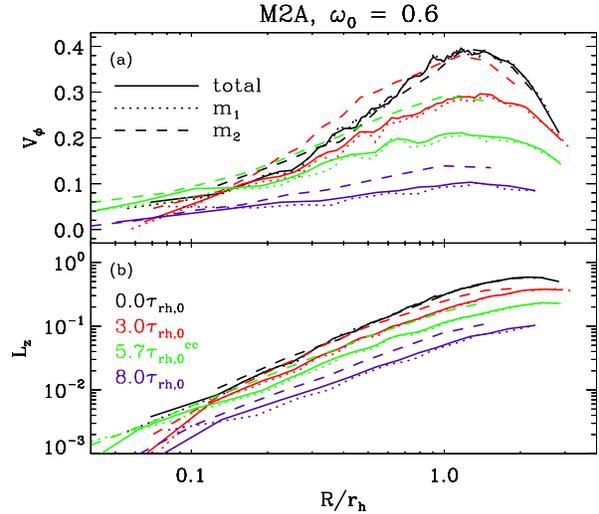}
  \caption{ Upper panel shows rotation curves at time \(T=0, 3, 5.7\) (core collapse) and \( 8\tau_{rh,0}\).
  Solid, dotted and dashed lines mean the rotation curves of all stars, high mass stars and low mass stars, respectively.
  At time \(T=0\), rotation curves are similar between low and high mass components.
  However, after few \(\tau_{rh,0}\), the rotation curve of \(m_{1}\) drops while that of \(m_{2}\) remains.
  The curve of z-direction angular momentum is shown in lower panel. Angular momenta also decrease with time. }
  \label{fig_rot06}
\end{figure}

\begin{figure}
  \centering
  \includegraphics[width=84mm]{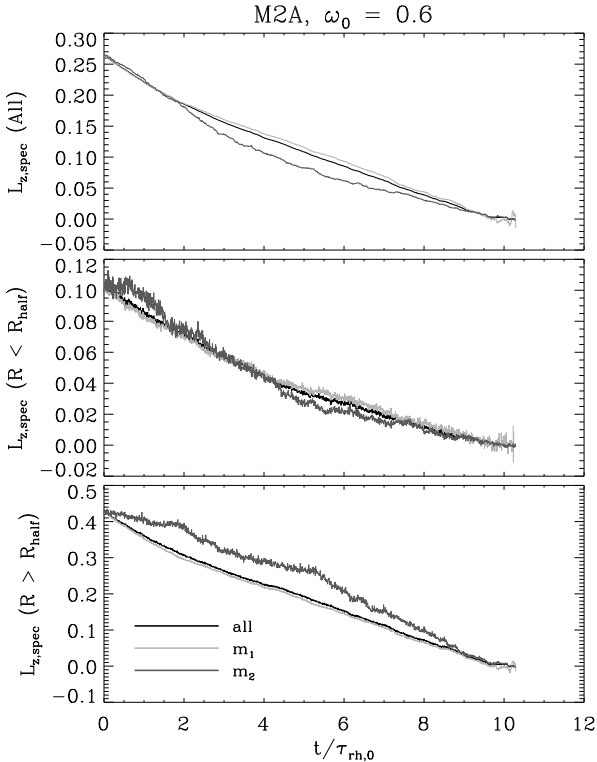}
  \caption{ Time evolution of z-direction specific angular momentum \(L_{z,spec}\) (upper), \(L_{z,spec}\) of inner half-mass (middle)
  and \(L_{z,spec}\) of outer half-mass (lower).
  %Black, red and blue lines represent the specific angular momentum for all, high mass and low mass stars, respectively.
  Though the angular momentum curve of high mass components is larger than
  low mass components at all radii, \(L_{z,spec}\) of high mass components in whole cluster becomes smaller than low mass components
  after few \(\tau_{rh,0}\). This is because most of high mass components are located in central region
  which rotates slowly due to the mass segregation. The evolution of the inner region shows similar behavior with that of whole cluster.
  On the other hand, for the outer region, \(L_{z,spec}\) of high mass components is higher than that of low mass components always.}
  \label{fig_ang3}
\end{figure}
To investigate the evolution of rotational properties of our cluster models, 
we focus on rotational velocities, \(V_{\phi}\) of two models in detail. 
As a representative of slowly rotating models, we chose a rotating model with \(\omega_{0} = 0.6\) of M2A.
Fig. \ref{fig_tr06} shows the distribution of \(V_{\phi}\) at four different epochs \(T=0, 3, 5.7\) and \(8\tau_{rh,0}\).
The core collapse time is \(T=5.7\tau_{rh,0}\). We combined results of three runs with \(N=\)20,000. 
At \(T=0\), the distribution is asymmetric to the positive direction due to the initial rotation.
The dispersion of \(V_{\phi}\) is large at the center and becomes smaller along the cylindrical radius, \(R\equiv (x^{2}+y^{2})^{1/2}\).
As the cluster loses its mass and angular momentum, 
the size of cluster and the degree of asymmetry in \(V_{\phi}\) decrease with time.
Finally, the distribution of \(V_{\phi}\) becomes symmetric compared to the initial distribution.
The rotation curves of both mass components are shown in Fig. \ref{fig_rot06}a.
The rotation curves of \(m_{1}\) and \(m_{2}\) are identical at \(T=0\)
because we assume that the distribution of positions and velocities for both mass components are the same initially. 
At \(T=3\tau_{rh,0}\), the rotation velocity of \(m_{1}\) becomes smaller at all radii
because low mass stars with large angular momenta escape from the system. 
On the other hand, the rotation curve of \(m_{2}\) remains as that of the initial curve for longer time.
At \(T=T_{cc}\), the rotation velocity of \(m_{2}\) also decreases but is still higher than that of \(m_{1}\).
The cluster rotates very slowly at \(T=8\tau_{rh,0}\). 
At this time, the rotation curves of \(m_{1}\) and \(m_{2}\) are flattened compared to other epochs.
It is interesting to note that the peak position remains at a constant value measured in the units of half-mass radius 
while the peak rotation velocity decreases.

Fig. \ref{fig_rot06}b shows the radial profiles of mean angular momentum for \(m_{1}\) and \(m_{2}\). 
The angular momentum curves of \(m_{1}\) and \(m_{2}\) are identical at \(T=0\) 
and \(L_{z}\) of \(m_{1}\) decreases more rapidly compared to that of \(m_{2}\).
Initially, the curve is a power-law with index of ~1.5 within \(r_{h}\) (e.g., the power law index is 2 for a rigid body rotation.).
The power law index becomes smaller and goes to one at \(T=8\tau_{rh,0}\) because the rotation curve becomes flatter.
In Fig. \ref{fig_ang3}, we show evolutions of specific angular momentum \(L_{z,spec}\) in different radial ranges:
in the whole cluster, within half-mass radius and outside of half-mass radius.
\(L_{z,spec}\) decreases with time due to the escaping stars with angular momenta.
In the entire cluster, \(L_{z,spec}\) of \(m_{1}\) and \(m_{2}\) decrease together until \(T=2\tau_{rh,0}\).
Although \(L_{z}\) of \(m_{2}\) is larger than that of \(m_{1}\) 
at the all radii as shown in Fig. \ref{fig_rot06}b, \(L_{z,spec}\) of \(m_{2}\) becomes smaller than that of \(m_{1}\) 
because high mass component tends to be more concentrated in the central region 
than low mass stars due to mass segregation (see \S 5.3 for more details).
Within the half-mass radius, \(L_{z,spec}\) of \(m_{1}\) decreases continuously 
while that of \(m_{2}\) remains at a constant value until \(T \approx 0.5 \tau_{rh,0}\)
because the rotation curve of high mass stars remains as the initial curve for a while as shown in Fig. \ref{fig_rot06}a
and Fig. \ref{fig_rot06}b.
After a few \(\tau_{rh,0}\), the evolution of inner region follows that of entire cluster due to the mass segregation.
For the outer region, \(L_{z,spec}\) of \(m_{2}\) is higher than that of \(m_{1}\) throughout the whole evolutionary phase.
Even for the mass segregation, the mean mass of outer region (see Fig \ref{fig_seg}) does not change significantly.
It means that the mass fraction of the outer region is less affected than that of inner region.
Thus, the mean rotation of \(m_{2}\) is still faster than the rotation of \(m_{1}\) beyond half-mass radius.
Finally, the cluster loses most of its mass and angular momentum at the end of evolution.

\subsubsection{Fast rotation (\(\omega_{0} = 1.5\))}
\begin{figure}
  \centering
 \includegraphics[width=84mm]{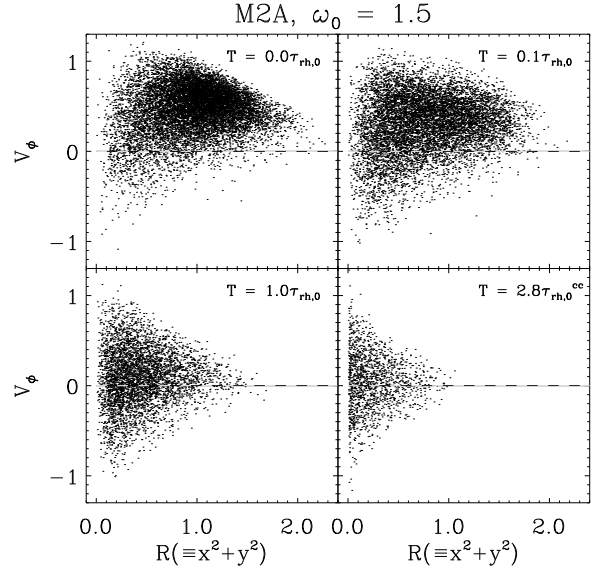}
  \caption{ Distribution of the rotational velocities of stars as a function of cylindrical radius for the model with \(\omega_{0} = 1.5\).
  Initially, most of stars have positive rotational velocities. However, the asymmetry rapidly decreases within a \(\tau_{rh,0}\).}
  \label{fig_tr15}
\end{figure}
\begin{figure}
  \centering
  \includegraphics[width=84mm]{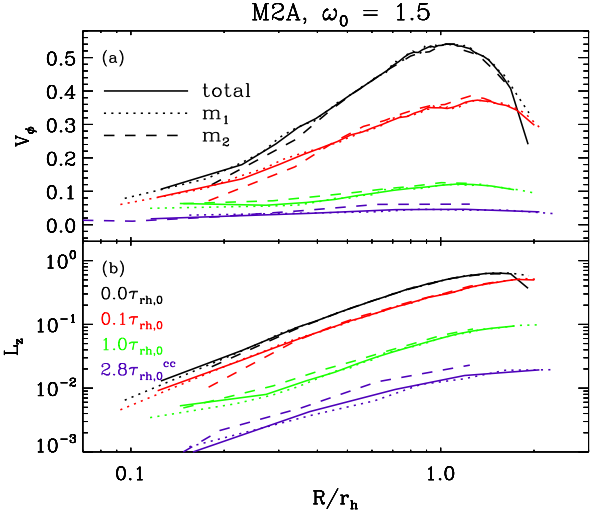}
  \caption{ Rotation curves at time  \(T = 0, 0.1, 1\) and 2.8 (core collapse) \(\tau_{rh,0}\) with \(\omega_{0}=1.5\) (upper).
  Lines have same meaning to those of Fig. \ref{fig_rot06}.
  At time \(T = 0\), rotation curves are similar between low and high mass components. Unlike rotation curves of in Fig \ref{fig_rot06},
  those of low and high mass components are not divided much. Curves of z-direction angular momentum are shown in lower panel.}
  \label{fig_rot15}
\end{figure}

 
We also investigate the evolution of rotational properties for the model with \(\omega_{0} = 1.5\).
As we mentioned earlier, this model is unstable against bar instability 
and the shape of cluster quickly becomes a prolate with the largest elongation at \(T\sim0.1\tau_{rh,0}\).
In Fig. \ref{fig_tr15}, we show the distribution of tangential velocities \(V_{\phi}\) 
of stars at \(T=0, 0.1, 1\) and \(2.8\tau_{rh,0}\) (core collapse).
Initially, the distribution is more skewed toward the positive direction than the model with \(\omega_{0} = 0.6\) as shown in Fig. \ref{fig_tr06}.
Only less than 10\% of stars have negative value at \(T=0\).
The distribution becomes symmetric and also the size of cluster becomes smaller as similar to the model with \(\omega_{0} = 0.6\).
At core collapse, only about 10\% of stars remain in the cluster and the cluster rotate slowly.
Fig. \ref{fig_rot15}a shows the rotation curve at \(T=0, 0.1, 1\) and \(2.8\tau_{rh,0}\).
The rotation curves of \(m_{1}\) and \(m_{2}\) decrease together,
indicating that the bar instability (i.e., large mass, energy and angular momentum loss) disturbs the relaxation processes.
The peak of rotational velocity decreases rapidly with time.
However, the peak position at \(T=0.1\tau_{rh,0}\) is slightly larger than the other epochs due to the effect of bar from the instability.
At core collapse, the rotation curve is nearly flat but still remains, though the cluster lose most of the mass and the angular momentum.
The mean angular momentum along the radius is shown in Fig. \ref{fig_rot15}b.
Similar to the result of model with \(\omega_{0} = 0.6\), it shows a power-law distribution.
Initially, the power law index is \(\sim\)2 within \(r_{h}\) like a rigid body rotation but it becomes close to 1 at core collapse. 

\section{Discussion}

\subsection{Mass evaporation}
\begin{figure}
  \centering
  \includegraphics[width=84mm]{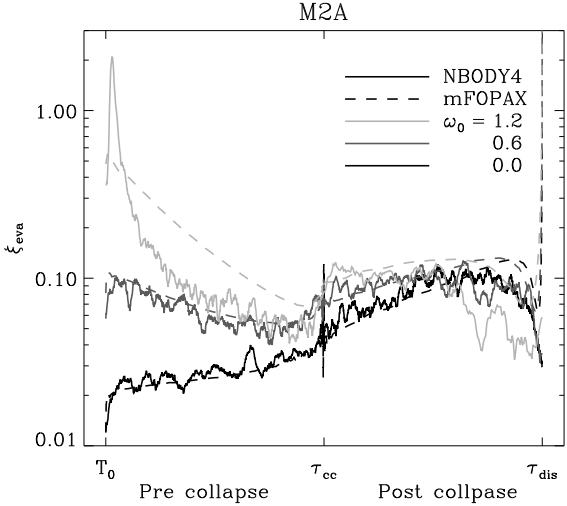}
  \caption{Dimensionless mass evaporation rate. X-axis means the evolutionary phase (i.e.; pre- or post-core collapse).
           For \(\omega_{0}=0.0\) and 0.6, the result of $N$-body and FP are well agreed. On the other hand,
           for \(\omega_{0}=1.2\), there is a spike meaning large mass loss induced by the bar instability in $N$-body result.}
  \label{fig_eva}
\end{figure}
To investigate the evolution of mass in detail, we first define the dimensionless mass evaporation rate such as
\begin{equation}
\xi_{e}\equiv-\frac{\tau_{rh}}{M}\frac{dM}{dt}
\end{equation}
where \(\tau_{rh}\) and \(M\) are the half mass relaxation time and total mass of a cluster at time \(T\), respectively.
Fig. \ref{fig_eva} shows the behavior of \(\xi_{e}\) of $N$-body and FP simulations
for M2A models with \(\omega_{0}=\)0.0, 0.6 and 1.2.
We divide the evolution into pre- and post-core collapse phases to investigate the evolution of \(\xi_{e}\) more clearly.
The mass evaporation rates are known to be constant for self-similar case
\citep[e.g.,][]{1961AnAp...24..369H, 1987ApJ...322..123L}.
However, the rate changes with time because our models are not self-similar.
In early phase, \(\xi_{e}\) increases with the initial rotation.
For slowly rotating models, we see a very good agreement between $N$-body and FP results in pre-core collapse phase.
On the other hand, for model with \(\omega_{0}=1.2\), there is a significant difference between results of $N$-body and FP.
The $N$-body results show a spike at the early time while FP results decrease monotonically. 
This spike is related to the rapid mass loss induced by the bar instability.
After the spike, \(\xi_{e}\) of $N$-body suddenly decreases below \(\xi_{e}\) of FP result.
It is interesting that \(\xi_{e,cc}\), the mass evaporation rate at the core collapse,
of $N$-body and FP show similar results even though there is a big difference before core collapse.
After core collapse, \(\xi_{e}\) of $N$-body and FP increase toward a peak value and decrease afterward.
The large differences of \(\xi_{e}\) between $N$-body and FP results at the end of the evolution is 
due to small number of remaining stars and thus do not have statistical significance.
\citet{2002MNRAS.334..310K} carried out FP simulations for rotating clusters with single mass system.
They also calculated \(\xi_{e}\) with different initial rotation \(\omega_{0}=\)0.0, 0.3 and 0.6.
We notice that the evolutionary shapes are similar between single mass and the 2-component mass systems.
However, \(\xi_{e}\) for 2-component mass systems show about 30\% enhancement
compared to \(\xi_{e}\) for single mass systems in pre-core collapse phase.
This enhancement could have been induced by the energy exchange process in multi-component models, 
as noticed by \citet{1995ApJ...443..109L}.
They calculated \(\xi_{e}\) with various initial mass functions and found that
\(\xi_{e}\) increases when the cut-off mass ratio (i.e., \(m_{f}/m_{i}\) if the mass ranges from \(m_{i}\) to \(m_{f}\)) increases.
To confirm the relationship between the mass evaporation rate and the mass ratio, 
we compute the maximum evaporation rate after the core collapse \(\xi_{e,post}\) for $N$-body results with various mass spectra. 
\begin{figure}
  \centering
  \includegraphics[width=84mm]{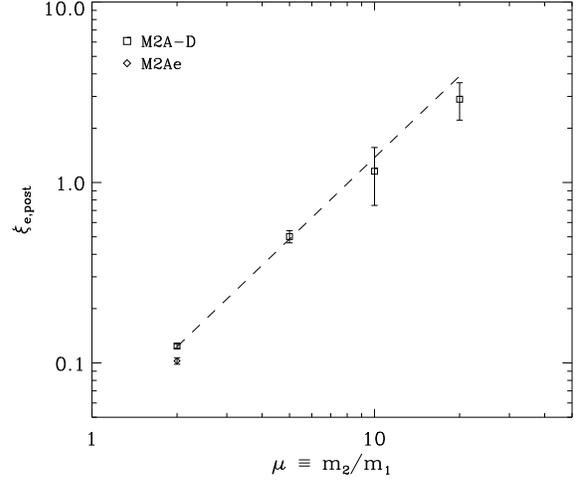}
  \caption{Maximum mass evaporation rate \(\xi_{e,post}\) after core collapse as a function of the individual mass ratio.
  Error bars show standard deviations of data of \(\xi_{e,post}\) for different mass spectra.}
  \label{fig_meva}
\end{figure}
Fig. \ref{fig_meva} shows \(\xi_{e,post}\) as a function of the individual mass ratio \(m_{2}/m_{1}\).
As shown in Fig. \ref{fig_eva}, the peak \(\xi_{e}\) in post-core collapse phase
are very similar with different initial rotations, so we average the results of each mass spectrum for different initial rotations.
\(\xi_{e,post}\) of M2A-D models increases with increasing \(m_{2}/m_{1}\) and follows a simple power-law.
For M2Ae model, \(\xi_{e,post}\) is slightly smaller than that of M2A model 
because the fraction of high mass stars for M2Ae model is small.

\subsection{Energy equipartition}
As presented in \S 4.2, high and low mass stars in the core approach to the `thermal' equilibrium state by the two body relaxation.
To investigate the energy equipartition in detail, we adopt the equipartition parameter
\begin{equation}
\xi_{eq}=\frac{m_{2}\sigma_{2}^{\phantom{0}2}}{m_{1}\sigma_{1}^{\phantom{0}2}}
\end{equation}
like previous studies \citep{2000ApJ...539..331W,2007MNRAS.374..703K}.
We calculated \(\xi_{eq}\) for stars inside the core radius. 
\begin{figure}
  \centering
  \includegraphics[width=84mm]{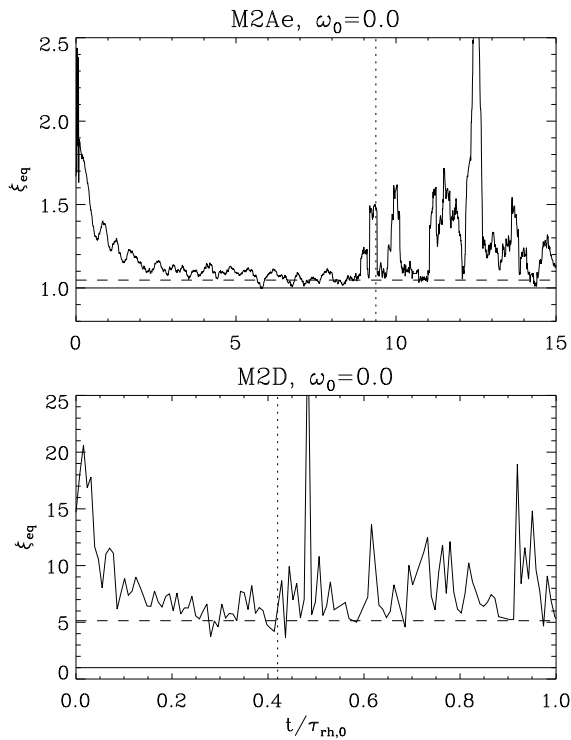}
  \caption{ Examples for the evolution of the equipartition parameter \(\xi_{eq}\).
  Dashed lines mean the minimum values \(\xi_{eq,min}\), and dotted lines show the core collapse time.
  The determination of \(\xi_{eq,min}\) is mentioned in text.}
  \label{fig_equ}
\end{figure}
Fig. \ref{fig_equ} shows the evolution of \(\xi_{eq}\) for models M2Ae and M2D without initial rotation for examples.
Dotted lines represent the core collapse time.
For the M2Ae model, \(\xi_{eq}\) approaches to unity and becomes less than unity at some moments.
However, for the M2D model, \(\xi_{eq}\) never approaches to unity.
The data is very noisy after the core collapse because only a few low mass stars remain in the core.
To determine the minimum value \(\xi_{eq,min}\), we find the best polynomial fitting function by varying the order from 5 to 15.

\begin{table*}
  \begin{center}
  \caption{\(\xi_{eq,min}\) of all models} 
  \begin{tabular} {c c c c c c c c c c c}
    \hline
    \hline
    Model&$S$& \(\Lambda\) & \(\omega_{0}\) & \(\xi_{eq,min}\) & & Model&$S$& \(\Lambda\) & \(\omega_{0}\) & \(\xi_{eq,min}\) \\
    \hline
M2A &0.566&1.056& 0.0 & 1.105 & &  M2C&6.325&50.24& 0.0 & 3.017 \\
            &-&-& 0.3 & 1.091 & &             &-&-& 0.6 & 2.923 \\
            &-&-& 0.6 & 1.107 & &             &-&-& 1.2 & 3.314 \\
            &-&-& 0.9 & 1.123 & & M2D&17.89&256.2& 0.0 & 5.141 \\
            &-&-& 1.2 & 1.158 & &             &-&-& 0.6 & 5.224 \\
            &-&-& 1.5 & 1.201 & &             &-&-& 1.2 & 6.672 \\
M2B &2.236&9.518& 0.0 & 1.664 & & M2Ae&0.141&0.264& 0.0 & 1.047 \\
            &-&-& 0.6 & 1.630 & &             &-&-& 0.6 & 1.071 \\
            &-&-& 1.2 & 1.860 & &             &-&-& 1.2 & 1.000 \\
    \hline
  \end{tabular}
  \label{tbl4}
  \end{center}
\end{table*}
Table \ref{tbl4} lists \(\xi_{eq,min}\) of all models. Only M2Ae models with \(\omega_{0}=0.0\) and 1.2 have lower \(\xi_{eq,min}\) than 1.05 
which is the value used in \citet{2007MNRAS.374..703K} for the energy equipartition.
On the other hand, \(\xi_{eq,min}\) increases when the equipartition instability parameters $S$ or \(\Lambda\) become larger.
\(\xi_{eq,min}\) for slowly rotating models (i.e., \(\omega_{0} \le 0.6\)) are similar to each other.
There are, however, significant differences of \(\xi_{eq,min}\) between slowly rotating models and rapidly rotating models.
For M2A models with \(\omega_{0} \ge 0.9\), \(\xi_{eq,min}\) increases gradually with initial rotation.
This is another phenomenon of the bar instability obstructing the relaxation process.

\subsection{Mass segregation}
\begin{figure}
  \centering
  \includegraphics[width=84mm]{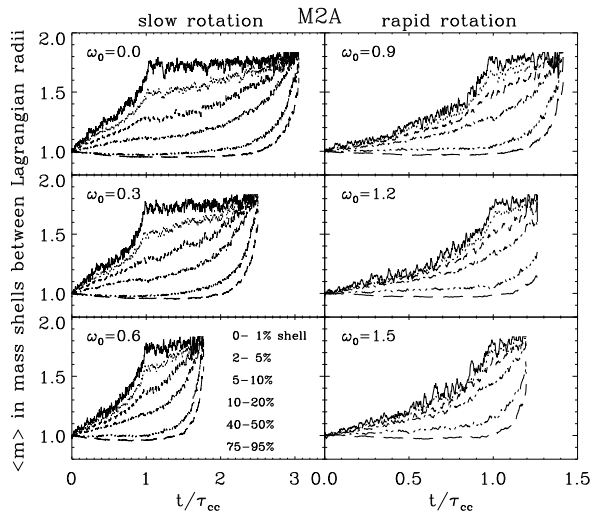}
  \caption{ Mean mass in mass shells between different Lagrangian radii with different initial rotations.
  Mean masses of inner shells increase with time before the core collapse while mean masses of outer shells decrease.
  This shows the process of the mass segregation. After the core collapse, mean masses of other shells also increase with somewhat time gaps
  but this is not from the mass segregation but because low mass stars escape more frequently than high mass stars.}
  \label{fig_seg}
\end{figure}
A simple method measuring the degree of the mass segregation is suggested by \citet{1996MNRAS.279.1037G}.
They calculated the mean mass in a space between different Lagrangian radii (i.e., Lagrangian shells) 
to see the change of mass distribution in un-equal mass systems.
More recently, \citet{2007MNRAS.374..703K} carried out $N$-body simulations with different mass spectra
and found that the mass segregation occurs inward direction (i.e., the mean mass of each shell is decoupled stepwise from outside, see models A and B in Fig. 6 of \citet{2007MNRAS.374..703K}).
Fig. \ref{fig_seg} shows the evolution of mean mass in different Lagrangian shells as a function of time.
In pre-core collapse phase, due to the mass segregation, the mean mass of the innermost shell increases while mean masses decrease in outer shells.
Note that the maximum mean mass of the innermost shell does not depend on the initial rotation
and also the innermost shell is nearly fully-occupied by high mass stars (i.e., \(<m>\approx m_{2}=1.833\)) at the time of core collapse.
The mean mass of the innermost shell stays at a constant value after the core collapse.
According to \citet{1996MNRAS.279.1037G}, after the core collapse, 
mean masses of inner shells with \(r<r_{20\%}\) slightly decrease because high mass stars are removed by binary formation 
and the mass distribution finally reaches a steady state in Lagrangian coordinate although the system expands with time.
Our simulations, however, show significant differences from the previous studies 
\citep{1996MNRAS.279.1037G, 2007MNRAS.374..703K} in post-core collapse phase.
In Fig. \ref{fig_seg}, mean masses of inner shells continue to increase even after the core collapse.
This is because our models are tidally-limited. 
Low mass stars escape from the cluster more rapidly than high mass stars as shown in Fig. \ref{fig_mss}.
As most of low mass stars escape, mean masses of outer shells increase at the end of evolution.

\subsection{Angular Momentum Exchange}
\begin{figure}
  \centering
  \includegraphics[width=84mm]{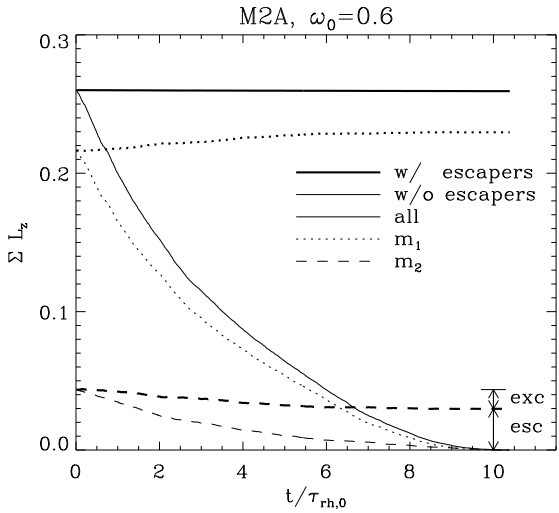}
  \caption{ Evolution of the angular momentum for the model with \(\omega_{0}=0.6\).
  When we consider the angular momentum of escapers, the total angular momentum is conserved (thick solid line).
  The total angular momentum of \(m_{1}\) slightly increases (thick dotted line),
  while that of \(m_{2}\) decreases (thick dashed line). This shows the existence of the angular momentum exchange from \(m_{2}\) to \(m_{1}\).
  The words `esc' and `exc' mean the degree of the angular momentum loss by escape and exchange, respectively.}
  \label{fig_ang060}
\end{figure}
Because our $N$-body model includes two different mass components, 
we are able to investigate the angular momentum exchange between different mass components.
This has not been studied carefully yet and is an important subject to understand the evolution of rotating star clusters.
However, it is not easy to analyze the angular momentum transfer with the tidal boundary 
because total angular momentum of cluster decreases continuously by escaping stars.
To distinguish the loss of angular momentum of a cluster between escaping and exchange, 
we register the positions and velocities of each escaping stars at the moment of escape.
Fig. \ref{fig_ang060} shows the time evolution of the total angular momentum for a model with \(\omega_{0}=0.6\).
Thin lines mean the total angular momentum of stars within the cluster and thick lines mean those of stars including escapers.
From the thick solid line, we observe that the total angular momentum including escapers is conserved as expected.
Interestingly, the angular momentum of \(m_{2}\) including escapers decreases while that of \(m_{1}\) increases.
Therefore, we conclude that there is a transfer of angular momentum from \(m_{2}\) to \(m_{1}\). 

\begin{figure}
  \centering
  \includegraphics[width=84mm]{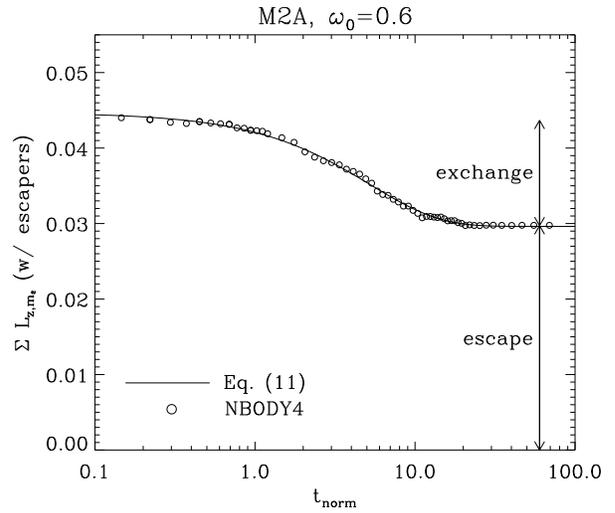}
  \caption{ Detailed evolution of the angular momentum of \(m_{2}\) including those of escapers (i.e., thick dashed line
  in Fig. \ref{fig_ang060}) as a function of \(t_{norm}\).
  \(t_{norm}\) is a time unit normalized by half-mass relaxation time (see the text).
  Open circles represent the $N$-body results.
  At the end of evolution, one can distinguish the amount of angular momentum loss by exchange and escape.
  The solid line shows the Eq. (11) with \(\xi_{exc}\) in Table \ref{tbl5}.
  Note that the line is shifted as much as the amount of angular momentum loss by escape.
  }
  \label{fig_exc060}
\end{figure}
If the exchange of angular momentum is due to the two-body relaxation, 
we can define the angular momentum exchange rate as follows
\begin{equation}
\xi_{exc}\equiv-\frac{\tau_{rh}(t)}{L_{z,m_{2}\to m_{1}}}\frac{dL_{z,m_{2}\to m_{1}}}{dt},
\end{equation}
where \(L_{z,m_{2}\to m_{1}}\) is the amount of remaining angular momentum expected to be transferred from \(m_{2}\) to \(m_{1}\).
If the angular momentum transfer rate is a constant, the above equation can be integrated to give,
\begin{equation}
L_{z,m_{2}\to m_{1}} = L_{z,m_{2}\to m_{1}}(0) e^{-\xi_{exc} t_{norm}}
\end{equation}
when we use a new time unit normalized by half-mass relaxation time \(t_{norm}=\int 1/\tau_{rh}(t) dt\).
This rate can be a good parameter to measure the efficiency of the angular momentum transfer between different mass components.
The angular momentum of clusters goes to 0 when clusters are disrupted. 
So we divide the loss of angular momentum of \(m_{2}\) into two contributions: through escaping and through transferring to \(m_{1}\).
We also estimate the fractional angular momentum loss by two different processes.
In Fig. \ref{fig_exc060}, the detailed evolution of the angular momentum of \(m_{2}\) including those of escapers is represented by open circles. 
At the end of evolution, whole amount of angular momentum of \(m_{2}\) disappears by escape or exchange.
The solid line which is from Eq. (11) with suitable value of \(\xi_{exc}\) 
and shifted as much as the amount of angular momentum loss by escape agrees well with the $N$-body result.
Table \ref{tbl5} shows the initial total angular momentum of \(m_{2}\), the fraction of angular momentum loss by escaping and exchanging 
and the angular momentum exchange rate.
Although the angular momentum exchange rate increases with the initial rotation, the fraction of angular momentum loss
by exchange decreases. For the model with \(\omega_{0}=1.5\), the fraction is less than 10\%. 
Rapidly rotating models have larger angular momentum exchange rate than slowly rotating models,
but their lifetimes are very short compared to slowly rotating models.
Therefore, these rapidly rotating models do not have enough time to exchange angular momentum from \(m_{2}\) to \(m_{1}\).
\begin{table}
  \begin{center}
  \caption{Angular momentum loss by escape and exchange for M2A models}
  \begin{tabular} {c c c c c c}
    \hline
    \hline
    \(\omega_{0}\) &  \(L_{z,0}(m_{2})\) & escape (\%) & exchange (\%) & \(\xi_{exc}\) \\
    \hline
    0.3 & 0.0283 & 57.8 & 42.2 & 0.17 \\ 
    0.6 & 0.0448 & 66.2 & 33.8 & 0.19 \\
    0.9 & 0.0555 & 70.4 & 29.6 & 0.23 \\
    1.2 & 0.0571 & 88.1 & 11.9 & 0.34 \\
    1.5 & 0.0635 & 91.3 & 8.7  & 0.71 \\
    \hline
  \end{tabular}
  \label{tbl5}
  \end{center}
\end{table}

\section{Summary and Conclusion}
We performed numerical simulations of rotating stellar system with two mass components using \textsc{nbody4}
and mFOPAX codes. By considering various mass spectra, 
we confirmed that both the initial rotation and the mass spectrum accelerate 
the evolution of the stellar system, as presented previous studies \citep{1999MNRAS.302...81E, 2002MNRAS.334..310K, 2004MNRAS.351..220K, 2008MNRAS.383....2K}.
However, we found that the initial rotation does not affect the evolution before the core collapse 
when the individual mass ratio \(m_2/m_1\) is large enough.
The mass evaporation rate is closely related to the acceleration of the evolution and increases with \(m_2/m_1\). 

According to the instability criteria from \citet{1973ApJ...186..467O}, we classified our models to slowly rotating models (i.e., \(T_{rot}/|W|<\)0.14)
and rapidly rotating models (i.e., \(T_{rot}/|W|>\)0.14).
By comparing the results of different approaches, $N$-body and FP simulations,
we confirmed that two approaches agree well with small differences on the time scales 
for the slowly rotating models.
On the other hand, for rapidly rotating models, there are significant discrepancies between $N$-body and FP results.
From the investigation of shape of systems, we revealed that the bar instability happens 
at the beginning for rapidly rotating models in $N$-body simulations.
This bar instability induces unexpected phenomena like the rapid loss of mass, energy and angular momentum. 
In addition, the bar instability hinders the two-body relaxation process, 
so the dynamical evolution of rapidly rotating systems is delayed as compared with FP results.
We therefore concluded that the 2 dimensional FP approach is not valid for rapidly rotating cases 
because 2 dimensional FP approaches are unable to treat non-axisymmetric models.

As the result of two-body interactions, low and high mass stars exchange their kinetic energies and happen to have similar kinetic energies. 
We confirmed that our models agree well with the equipartition instability criteria \citep{1969ApJ...158L.139S,2000ApJ...539..331W} for slowly rotating models. 
When the mass ratio becomes larger, it is hard to reach the complete equipartition state.
Moreover, the equipartiton process is more disturbed for rapidly rotating models which suffer the bar instability.
We also observed the exchange of angular momentum between low and high mass stars by investigating escapers 
and defined the angular momentum exchange rate \(\xi_{exc}\). \(\xi_{exc}\) increases when the initial rotation increases.
However, the amount of transferred angular momentum from high mass stars to low mass stars 
decrease because clusters with rapid initial rotation survive rather shortly compared to those with slow initial rotation.

\section*{Acknowledgement}
We would like to thank to Sverre Aaseth and Junichiro Makino who have developed excellent software and hardware. 
We also thank Andreas Ernst for reading the manuscript. 
This research was supported by the KRF grant No. 2006-341-
C00018. The computation was done on the GPU computer provided
by a grant from the National Institute for Mathematical Sciences
through the Engineering Analysis Software Development program.
RS acknowledges support through the Silk Road Project and
Chinese Academy of Sciences Visiting Professorship for Senior
International Scientists, Grant Number $2009S1-5$.

\label{lastpage}
\end{document}